\documentclass[10pt,A4paper]{article}
\usepackage{geometry}
\usepackage{amsfonts}
\usepackage{amsmath}
\usepackage{amssymb}
\usepackage{mathrsfs}
\usepackage{graphicx}
\usepackage{hyperref}
\usepackage[usenames,dvipsnames]{color}
\usepackage{cancel}
\usepackage{cleveref}
\usepackage{bbm}
\usepackage{bm}
\usepackage{mathtools}
\usepackage{physics}
\usepackage{tensor}
\usepackage{array}
\usepackage{float}
\usepackage{wrapfig}
\usepackage{soul}
\usepackage{xcolor}
\usepackage{cite}
\usepackage{url}

\hypersetup{
colorlinks=true,
linkcolor=blue,
citecolor=red,
urlcolor=cyan
}

\title{\vspace{-1cm} \bf {\Large Total momentum and other Noether charges \\for particles interacting in a quantum spacetime}}
\date{}
\author{{Giovanni Amelino-Camelia, Giuseppe Fabiano, and {Domenico Frattulillo}}
\vspace{12pt}
\\
\small Dipartimento di Fisica Ettore Pancini, Universit\`a di Napoli ``Federico II'';
\\
\small and INFN, Sezione di Napoli, Complesso Univ. Monte S. Angelo, I-80126 Napoli, Italy;
}

\begin{document}
\maketitle
\begin{abstract}
    There has been  strong interest in the fate of relativistic symmetries
in some quantum spacetimes, also because of its possible relevance for high-precision experimental tests
of relativistic properties. However, the main technical results
obtained so far concern the description of suitably deformed relativistic-symmetry transformation rules,
whereas the properties of the associated Noether charges,
which are crucial for the phenomenology,
are still poorly understood.
We here tackle this problem focusing on first-quantized particles described within a Hamiltonian framework
and using as toy model the so-called ``spatial kappa-Minkowski noncommutative spacetime'', where
all the relevant conceptual challenges are present but, as here shown, in technically manageable fashion.
We derive the Noether charges, including the much-debated total-momentum charges,
and we expose a strong link between the properties
of these Noether charges and the structure of the laws of interaction among particles.
\end{abstract}

\section{Introduction}
The structure of the quantum-gravity problem invites us to contemplate the possibility that spacetime itself might be affected by one form or another of quantization \cite{Snyder:1946qz,Seiberg:1999vs,Doplicher:1994tu,Amelino-Camelia:2016gfx,Oriti:2006se,tHooft:1996ziz}. Besides (some remnant of) the light-cone structure specified by the speed-of-light scale, a quantum spacetime would inevitably also have
additional structure, concerning its quantization, which in most models is specified in terms of a length scale. There has been strong interest \cite{Amelino-Camelia:1997ieq,Amelino-Camelia:2011lvm, Smolin:2005cz,Girelli:2004ms,Gambini:1998it,Alfaro:1999wd} in understanding if and how this additional structure could affect some relevant relativistic properties, an issue which, besides its conceptual appeal,
might also bear some relevance for phenomenology, since some relativistic properties can be tested with very high accuracy. Among the most studied models that fit this profile there are
some noncommutative spacetimes which are known to require a deformation of relativistic symmetries such that
the noncommutativity length scale plays the role of second relativistically-invariant scale \cite{Amelino-Camelia:2002uql,Kowalski-Glikman:2002iba,Magueijo:2002am},
 in addition to the speed-of-light scale, but the phenomenology of this scenario has been stagnating because,
 lacking a full generalization of the Noether theorem\footnote{It has been
 established \cite{Amelino-Camelia:2007yca,Amelino-Camelia:2007une} that there is a generalization of the Noether theorem
 which applies to free theories formulated in some noncommutative spacetimes; however, a free theory has
 no phenomenology and it cannot even provide intuition for how the charges of different particles should be
 combined in conservation laws relevant for particle reactions.}, one is unable to reliably identify
 the relativistic-symmetry conserved charges, whereas of course the conserved charges play
 a key role in the experimental tests of relativistic symmetries. There have been attempts to
 formulate tentatively a phenomenology of these deformed relativistic symmetries by trying to
 guess how the charges of different particles should be
 combined in conservation laws relevant for particle reactions, with the guessing inspired by one or
 another ``naturalness argument" based on the form of the relativistic properties of free particles,
 but it was already clear (and will be here confirmed) that the only reliable path requires
 a constructive derivation of conserved charges in interacting theory.

While previous attempts all focused on action/Lagrangian formulations of theories
in noncommutative spacetimes \cite{Bevilacqua:2022fbz,Douglas:2001ba,Amelino-Camelia:2007yca,Amelino-Camelia:2007une},
we here investigate this Noether-charges issue within a Hamiltonian setup, finding
that this provides several advantages.
As illustrative example of Noether-charges analysis in a quantum spacetime we focus on
the so-called ``spatial kappa-Minkowski noncommutative spacetime", where
all the relevant conceptual challenges are present but, as here shown, in technically manageable fashion.
We exhibit some examples of Hamiltonians describing two-particle and three-particle interactions for which
the Noether charges can be constructively derived.
A key bring-home message is that within a given description of (deformed) relativistic symmetries for
the free-particle case, the total charges that are then conserved when one allows multiparticle interactions
depend strongly on the form of the Hamiltonian, also exposing the
weakness of
previous ``naturalness arguments" used for guessing the Noether charges.

\section{Preliminaries}
\label{prel}

Before getting to our novel results, we devote this section to a short review of properties of
spatial 2D $\kappa$-Minkowski
and to a general perspective on possible noncommutative-spacetime generalizations of harmonic-oscillator-type Hamiltonians (the type of Hamiltonians for which, in the following sections, we shall derive Noether charges).

\subsection{Spatial 2D $\kappa$-Minkowski}
The most studied variant of $\kappa$-Minkowski
noncommutativity is a case of space/time noncommutativity (spatial coordinates commute among themselves but do not commute with the time coordinate), which in the 2D case is characterized by the following commutator between time and spatial coordinate \cite{Majid:1994cy}
\begin{equation}
\label{spacetimekappa}
     [x^0,x^1]=i\ell x^1
\end{equation}
where $\ell$ (often rewritten as $1/\kappa$) is a length scale usually assumed to be of the order of the Planck length.
It is well established \cite{Majid:1994cy,Arzano:2021scz,Lukierski:1992dt}
that the symmetries of 2D
space/time $\kappa$-Minkowski
noncommutativity
are described by the 2D $\kappa$-Poincaré Hopf Algebra.

In the study we are here reporting we follow
Ref.\cite{Amelino-Camelia:2014gga} by focusing on a scenario with a time coordinate which is fully commutative and two spatial coordinates governed by
$\kappa$-Minkowski
noncommutativity
\begin{equation}
\label{spatialkappa}
    [x_2,x_1]=i\ell x_1 \, .
\end{equation}
All the results establishd in a wide literature on the 2D space/time  $\kappa$-Minkowski of Eq.(\ref{spacetimekappa})
and its Hopf-algebra symmetries are easily converted into results for  our 2D spatial $\kappa$-Minkowski
of Eq.(\ref{spatialkappa})
and its Hopf-algebra symmetries, by the replacement of coordinates
$x^0\rightarrow ix_2$, a replacement of noncommutativity parameter
$\ell\rightarrow i\ell$, and then replacing the time-translator generator with a suitable generator of translations along the $x_2$
direction,
$P_0\rightarrow -iP_2$
while the boost generator of
2D space/time  $\kappa$-Minkowski
is replaced by the rotation generator of 2D spatial  $\kappa$-Minkowski,
$N\rightarrow -iR$. This leads to a description of the translation and rotation symmetries of 2D spatial  $\kappa$-Minkowski such that
\begin{equation}
\label{spatialalg}
    [P_2,P_1]=0\qquad [R,P_2]=-i P_1\qquad [R,P_1]=\frac{i}{2\ell}(1-e^{-2\ell P_2})+i\frac{\ell}{2}P_1^2
\end{equation}
which is a deformation of the Euclidean algebra in 2 dimensions. A central element of this algebra, which will be a crucial ingredient for the construction of our Hamiltonians, is given by
\begin{equation}
\label{cas}
    \mathcal{C}=\frac{4}{\ell^2}\sinh^2({\ell P_2/2})+e^{\ell P_2}P_1^2
\end{equation}
This is a deformation of the $P_1^2+P_2^2$ Casimir element of the Euclidean algebra.

We shall introduce interactions among particles within a Hamiltonian setup and be satisfied showing our results to order $\ell^2$. We note here some commutation relations
which shall be valuable in those Hamiltonian analyses:
\begin{equation}
    [x_1,P_1]=i  \quad
    [x_1,P_2]=0 \quad
    [x_2,P_1]=-i\ell P_1 \quad
    [x_2,P_2]=i
\end{equation}
\begin{equation}
\begin{aligned}
   & [R,x_1]=i x_2 \\
   &[R,x_2]=-i\big(x_1-\ell x_1P_2+\frac{\ell}{2} x_2P_1+\frac{\ell}{2} P_1x_2+\ell^2 x_1 P_2^2 +\frac{\ell^2}{4}(x_1 P_1^2+P_1^2 x_1)\big)
    \end{aligned}
\end{equation}
which satisfy Jacobi identities.

The nonlinearity of the commutators
(\ref{spatialalg}), typical of Hopf-algebra symmetries,
produce the difficulties for Noether charges which are the main focus of the study we are here reporting. For free particles it has been shown \cite{Amelino-Camelia:2007yca,Amelino-Camelia:2007une} that the charges associated to
$P_1$, $P_2$ and $R$ are conserved (but of course any nonlinear function of a conserved quantity is also conserved).
For interacting particles it is unclear which combinations of the charges should be conserved in particle reactions. In particular, for a process $A+B \rightarrow C+D$ it is clear that $P_1^A + P_1^B = P_1^C + P_1^D$ is not an acceptable conservation law because of the nonlinearity of $[R,P_1]$ ({\it i.e.} $P_1^A + P_1^B = P_1^C + P_1^D$ would not be covariant). So it is clear that the total momentum of a system composed of particles $A$ and $B$ cannot have the component $P_1^A + P_1^B$, but it is not clear which nonlinear combination of the momenta gives the total momentum of a system (and would be therefore conserved in particle reactions). A popular way to guess the momentum-composition formula is based on the so-called ``coproduct" \cite{Agostini:2002de,Kosinski:1999dw,Majid:1999tc}, which for our purposes it is sufficient to introduce in terms of the properties of suitably ordered products of plane waves: for two plane waves of momenta $k$ and $q$ one has that, as a result of
the noncommutativity (\ref{spatialkappa}),
\begin{equation}
e^{ik_1x^1}e^{ik_2x^2}e^{iq_1x^1}e^{iq_2x^2}=e^{i(k\oplus_\kappa q)_1x^1}e^{i(k\oplus_\kappa q)_2x^2}
\end{equation}
where
\begin{equation}
\begin{aligned}
\label{opm}
        &(k\oplus_\kappa q)_1=k_1+e^{-\ell k_2}q_1\\
        &(k\oplus_\kappa q)_2=k_2+q_2
\end{aligned}
\end{equation}
 In order for these quantities to close the single-particle algebra \eqref{spatialalg}, rotations should also combine non-linearly
\begin{equation}
\label{opr}
    (R_k\oplus_\kappa R_q)=R_k+e^{-\ell k_2}R_q
\end{equation}

Alternative ways for guessing the momentum-composition formula have also been proposed. As an alternative to
the ``$\kappa$-coproduct composition law"
of Eqs.(\ref{opm})-(\ref{opr})
we shall also consider the ``proper-dS composition law",
\begin{equation}
\label{dschargeextra}
\begin{aligned}
\mathcal{P}_1=(p^A\oplus_{dS} p^B)_1&=p^A_1+p^B_1-\ell(p_2^A p_1^B+p_1^A p_2^B)+\\ &+\frac{\ell^2}{2}\big[(p_2^A p_1^B+p_1^A p_2^B)(p_2^A+p_2^B)-p_1^A(p_1^B)^2-(p_1^A)^2p_1^B\big]\\
\mathcal{P}_2=(p^A\oplus_{dS} p^B)_2&=p^A_2+p^B_2+\ell p_1^A p_1^B-\frac{\ell^2}{2}\big[-p_1^Bp_1^A(p_2^B+p_2^A) +p_2^A(p_1^B)^2+(p_1^A)^2p_2^B\big]\\
\mathcal{R}=(R^A\oplus_{dS} R^B)&=R^A+R^B
\end{aligned}
\end{equation}
which was motivated using some geometric arguments
(one can show that with these choices of composition laws momentum space acquires the geometrical structure of de Sitter space \cite{Amelino-Camelia:2013sba}).

\subsection{Deformations of harmonic-oscillator Hamiltonians}\label{HO}

Our next task is to introduce the class of Hamiltonians on which we shall focus our search of Noether charges. Their core ingredient is
the harmonic oscillator potential in $2$ spatial dimensions. We shall consider deformations of the Hamiltonian
\begin{equation}
\label{h0ab}
H^{AB}_0=\frac{(\Vec{p}^A)^2}{2m}+\frac{(\Vec{p}^B)^2}{2m}+\frac{1}{2}g(\Vec{q}^A-\Vec{q}^B)^2
\end{equation}
where $g$ is the coupling constant, the labels $A$ and $B$ refer to the two particles interacting, $\Vec{q}^J$ ($J \in \{A,B \}$) are ordinary commutative spatial coordinates, and
$\Vec{p}^J$ are the corresponding momenta, with standard Heisenberg commutators ($[q^J_j,p^K_k]=i\,\delta^{JK}\delta_{jk}$, with $J,K\in \{A,B\}$ and $j,k=1,2$). The total momentum and total angular momentum defined through
\begin{equation}
\Vec{P}=\Vec{p}^A+\Vec{p}^B \qquad R_0=R^A_0+R^B_0
\end{equation}
are conserved charges since they commute with the Hamiltonian, $[H^{AB}_0,\Vec{P}]=0$ and $[H^{AB}_0,R_0]=0$. Both the total generators $\{P_i,R_0\}$ and the single particle generators $\{p_i^I,R_0^I\}$ close the un-deformed Galilean algebra.

For reasons which shall soon be clear, we want to test
our approach also for interactions among more than two particles, and for that purpose our starting point is the
3-particle Hamiltonian
\begin{equation}
\label{trivialthree}
H^{ABC}_0=\frac{(\Vec{p}^A)^2}{2m}+\frac{(\Vec{p}^B)^2}{2m}+\frac{(\Vec{p}^C)^2}{2m}+\frac{1}{2}g(\Vec{q}^A-\Vec{q}^B)^2+\frac{1}{2}g(\Vec{q}^A-\Vec{q}^C)^2+\frac{1}{2}g(\Vec{q}^B-\Vec{q}^C)^2
\end{equation}
This is of interest to us particularly because the interacting potential $V_3(\Vec{q}^A,\Vec{q}^B,\Vec{q}^C)$ can be split into the sum $V_2(\Vec{q}^A,\Vec{q}^B)+V_2(\Vec{q}^A,\Vec{q}^C)+V_2(\Vec{q}^B,\Vec{q}^C)$ with $V_2$ having the same functional form for each pair of particles: in the case studies for which we perfomed our Noether-charge analyses this property cannot be maintained in presence of noncommutativity of coordinates.

Evidently, the Hamiltonian \eqref{trivialthree} commutes with the total charges defined as $\Vec{P}=\Vec{p}^A+\Vec{p}^B+\Vec{p}^C$ and $R_0=R^A_0+R^B_0+R^C_0$.

A key ingredient of our deformed Hamiltonians will be of course the kinetic term, for which we adopt the form
\begin{equation}
\label{ke}
H_K\equiv\frac{\mathcal{C}}{2m}\approx\frac{p_1^2}{2m}+\frac{p_2^2}{2m}+\ell\frac{p_1^2 p_2}{2m}+\ell^2\frac{p_1^2p_2^2}{4m}+\ell^2\frac{p_2^4}{24m}
\end{equation}
obtained from the Casimir element $\mathcal{C}$ of our Eq.(\ref{spatialalg}) (to order $\ell^2$).

We will look for suitable interaction potentials within some rather broad parametrizations. We parametrize the two-particle case as follows
\begin{equation}
\label{ansatz}
V^{AB}=V(\Vec{x}^A,\Vec{x}^B)=\frac{1}{2} g (\Vec{x}^A-\Vec{x}^B)^2+\ell g\sum\alpha_{ijk}^{IJK}p_i^Ix_j^Jx_k^K+\ell^2g\sum\beta_{ijkh}^{IJKH}p_i^Ip_j^Jx_k^Kx_h^H
\end{equation}
where $\alpha_{ijk}^{IJK}$ and $\beta_{ijkh}^{IJKH}$ are numerical coefficients and the sum extends both to spatial indices (lower case letters) and particle indices (upper case letters).

Similarly, for the three-particle case our {\it ansatz}
is given by
\begin{equation}
\label{ansatz3}
\begin{aligned}
V^{ABC}=V(\Vec{x}^A,\Vec{x}^B,\Vec{x}^C)=&\frac{1}{2} g (\Vec{x}^A-\Vec{x}^B)^2+\frac{1}{2} g (\Vec{x}^B-\Vec{x}^C)^2+\frac{1}{2} g (\Vec{x}^C-\Vec{x}^A)^2+\\
+&\ell g\sum\Tilde{\alpha}_{ijk}^{IJK}p_i^Ix_j^Jx_k^K+\ell^2g\sum\Tilde{\beta}_{ijkh}^{IJKH}p_i^Ip_j^Jx_k^Kx_h^H
\end{aligned}
\end{equation}
where $\Tilde{\alpha}_{ijk}^{IJK}$
and $\Tilde{\beta}_{ijkh}^{IJKH}$ are other sets of numerical coefficients and the particle indices run over $\{A,B,C\}$.

\section{Charges with proper-dS composition}
\label{pdS}
The debate on the alternative ways to combine charges in a $\kappa$-Minkowski setup has mainly relied on naturalness arguments based on the properties of free particles in $\kappa$-Minkowski. As announced in our opening remarks, we here intend to show that there is no notion of ``naturalness" at stake here:
how charges should combine depends on the form of the laws of interaction among particles (so evidently goes beyond the scopes of the description of free particles) and different composition laws can emerge from different descriptions of the interactions. We shall establish our case relying on Hamiltonian theories within first-quantized quantum mechanics, where the relevant issues can be seen in particularly vivid fashion.

We choose as our first task the one of exhibiting a Hamiltonian (within first-quantized quantum mechanics) which selects uniquely the proper-dS composition law,
which we already reviewed in Eq.(\ref{dschargeextra}) and we show again here for convenience:
\begin{equation}
\label{dscharge}
\begin{aligned}
\mathcal{P}_1=(p^A\oplus_{dS} p^B)_1&=p^A_1+p^B_1-\ell(p_2^A p_1^B+p_1^A p_2^B)+\\&+\frac{\ell^2}{2}\big[(p_2^A p_1^B+p_1^A p_2^B)(p_2^A+p_2^B)-p_1^A(p_1^B)^2-(p_1^A)^2p_1^B\big]\\
\mathcal{P}_2=(p^A\oplus_{dS} p^B)_2&=p^A_2+p^B_2+\ell p_1^A p_1^B-\frac{\ell^2}{2}\big[-p_1^Bp_1^A(p_2^B+p_2^A) +p_2^A(p_1^B)^2+(p_1^A)^2p_2^B\big]\\
\mathcal{R}=(R^A\oplus_{dS} R^B)&=R^A+R^B
\end{aligned}
\end{equation}
One can easily verify that $\mathcal{P}_1\,,\,\mathcal{P}_2\,,\,\mathcal{R}$ close the
algebra \eqref{spatialalg} up to order $\ell^2$, which we also rewrite here for convenience
\begin{equation}
[\mathcal{P}_2,\mathcal{P}_1]=0\qquad [\mathcal{R},\mathcal{P}_2]=-i\mathcal{P}_1\qquad [\mathcal{R},\mathcal{P}_1]=i(\mathcal{P}_2-\ell \mathcal{P}_2^2+\frac{\ell}{2}\mathcal{P}_1^2+\frac{2\ell^2 \mathcal{P}_2^3}{3})
\end{equation}
We start by showing that for the case of two particles interacting  there is a Hamiltonian
$H^{AB}_{dS}$, deformation of the  $H^{AB}_0$
of Eq.(\ref{h0ab}), such that $[\Vec{\mathcal{P}},
H^{AB}_{dS}]=0$ and $[\mathcal{R},
H^{AB}_{dS}]=0$.
As anticipated in Sub\cref{HO}, our Hamiltonian $H^{AB}_{dS}$
will be of the form
\begin{equation}
    H^{AB}_{dS} =
H^A_K + H^B_K + V^{AB}_{dS}
\end{equation}
where $H_K$ is fixed to be the one of
Eq.(\ref{ke}),
while $V^{AB}_{dS}$
must be specified consistently with Eq.(\ref{ansatz}), for some choice of the parameters that Eq.(\ref{ansatz}) leaves to be determined.

We work partly by reverse engineering: we use
$[\Vec{\mathcal{P}},
H^{AB}_{dS}]=0$ and $[\mathcal{R},
H^{AB}_{dS}]=0$ as conditions that must be satisfied by
the parameters
 of Eq.(\ref{ansatz}),
 and then, once we have such an acceptable $V^{AB}_{dS}$, we show that the resulting Hamiltonian $H^{AB}_{dS}$ {\underline{uniquely}} selects the proper-dS charges (\ref{dscharge})
 as its conserved charges.

We find that in particular the following choice of $V^{AB}_{dS}$
\begin{equation}
\begin{aligned}
\label{pdSsep}
    V^{AB}_{dS}=&\frac{g}{2}\Big[(\Vec{x}^A-\Vec{x}^B)^2+\\
    +&2\ell\Big(-p_2^A(x_1^A)^2+\frac{1}{2}p_1^Ax_1^Ax_2^A+\frac{1}{2}x_2^Ax_1^Ap_1^A+p_2^Ax_1^Ax_1^B-x_2^Ap_1^Ax_1^B +(A\leftrightarrow B)\Big)+\\
    +&\frac{1}{2}\ell^2\Bigl((p_1^B)^2(-2(x_2^A)^2+6x_2^Ax_2^B-2(x_2^B)^2)+4p_1^Bp_2^Bx_1^Ax_2^A-p_1^Ap_1^Bx_2^Ax_2^B+p_1^Bp_2^Ax_1^Ax_2^B+\\
    -&6p_1^Bx_1^Ax_2^Bp_2^B  -2p_1^Bx_1^B(p_2^Ax_2^A-x_2^Bp_2^B)-2(p_2^B)^2((x_1^A)^2-x_1^Ax_1^B-(x_1^B)^2)+\\
    +&p_2^Bp_1^Ax_2^Ax_1^B-2p_2^Bp_2^Ax_1^Ax_1^B-3p_2^Bx_2^Ax_1^Bp_1^B+2x_1^Ap_1^A(p_1^Ax_1^A-p_1^Ax_1^B+p_2^Ax_2^A+\\
    -&\frac{3}{2}p_2^Ax_2^B+\frac{3}{2}x_2^Bp_2^B)
    +x_2^Ap_2^Ax_1^Bp_1^B+(A\leftrightarrow B)\Bigr)\Big]
\end{aligned}
\end{equation}
is indeed such that
$[\Vec{\mathcal{P}},
H^A_K + H^B_K + V^{AB}_{dS}]=0$ and $[\mathcal{R},
H^A_K + H^B_K + V^{AB}_{dS}]=0$.

We observe that our
$V^{AB}_{dS}$
is symmetric under  exchange of the particles
(this is not always the case, see later). Most importantly, we find that indeed the
Hamiltonian
$H^A_K + H^B_K + V^{AB}_{dS}$
{\underline{uniquely}} selects the proper-dS charges (\ref{dscharge})
 as its conserved charges.
In order to see this we start from a general parametrization of the two-particle charges
\begin{equation}
\begin{aligned}
\label{chargeansatz}
   & P_1^{tot}=\sum p_1^I+\ell  \gamma_{i j}^{I J} p_i^I p_j^J + \ell^2 \Gamma_{i j k}^{I J K} p_i^I p_j^J p_k^K\\
   &     P_2^{tot}=\sum p_2^I+\ell  \theta_{i j}^{I J} p_i^I p_j^J + \ell^2  \Theta_{i j k}^{I J K} p_i^I p_j^J p_k^K\\
      &  R^{tot} = \sum R^I+\ell  \phi_{i}^{I J} p_i^I R^J + \ell^2  \Phi_{i j }^{I J K} p_i^I p_j^J R^K
\end{aligned}
\end{equation}
where $\gamma,\theta,\phi,\Gamma,\Theta,\Phi$ are sets of real coefficients and the sum is intended over particle indices ${I,J,K}$ (which take values in $\{A,B\}$)
and over the spatial indices ${i,j,k}$. We also require that no terms with all particle indices equal to each other are present, so that we recover the definition of single particle charge when the charges of the other particles are zero.

By requesting that these charges  commute with $H^A_K + H^B_K + V^{AB}_{dS}$
the parameters in Eq.\eqref{chargeansatz} get fully fixed, giving indeed
the proper-dS charges (\ref{dscharge}).

Next we turn to the corresponding three-particle case, for which the proper-dS composition
leads to the following formulas for the charges:
\begin{equation}
\label{dscharge3}
\begin{aligned}
\tilde{\mathcal{P}_1}&=((p^A\oplus_{dS} p^B)\oplus_{dS}p^C)_1=p^A_1+p^B_1+p^C_1-\ell\bigl(p_2^B(p_1^C+p_1^A)+p_2^C(p_1^B+p_1^A)+\\&+p_2^A(p_1^C+p_1^B)\bigr)+\frac{\ell^2}{2}\bigl(-2p_1^Ap_1^Bp_1^C-(p_1^B)^2p_1^A+2p_1^Ap_2^Bp_2^C-p_1^B(p_1^A)^2-(p_1^C)^2(p_1^B+p_1^A)+\\&-(p_1^C)(p_1^B+p_1^A)^2+2p_2^Cp_1^Bp_2^A+(p_2^B+p_2^A)(p_2^Bp_1^A+p_2^Ap_1^B)+\\&+(p_2^A+p_2^B+p_2^C)(p_2^C(p_1^B+p_1^A)+p_1^C(p_2^A+p_2^B))\bigr)\\\\
\tilde{\mathcal{P}_2}&=((p^A\oplus_{dS} p^B)\oplus_{dS}p^C)_2=p_2^A+p_2^B+p_2^C+\ell\qty(p_1^Bp_1^A+p_1^Cp_1^B+p_1^Ap_1^C)+\\
&-\frac{\ell^2}{2}\bigl(p_2^C(p_1^A+p_1^B)^2-p_1^C(p_2^C(p_1^B+p_1^A)+(p_1^B-p_1^A)(p_2^B-p_2^A))+\\
&+(p_1^C)^2(p_2^B+p_2^A)+(p_1^B-p_1^A)(-p_2^Bp_1^A+p_1^Bp_2^A)\bigr)\\ \\
\tilde{\mathcal{R}}&=(R^A\oplus_{dS} R^B)\oplus_{dS}R^C=R^A+R^B+R^C
\end{aligned}
\end{equation}
Evidently we must find a Hamiltonian
$H^{ABC}_{dS}$, deformation of the  $H^{ABC}_0$ of
Eq.(\ref{trivialthree}), such that $[\Vec{\tilde{\mathcal{P}}},
H^{ABC}_{dS}]=0$ and $[\tilde{\mathcal{R}},
H^{ABC}_{dS}]=0$.
As anticipated in Sub\cref{HO}, our Hamiltonian $H^{ABC}_{dS}$
will be of the form
\begin{equation}\label{Hds3}
H^{ABC}_{dS} =
H^A_K + H^B_K+ H^C_K + V^{ABC}_{dS}
\end{equation}
where $H_K$ is again fixed to be the one of
Eq.(\ref{ke}),
while $V^{ABC}_{dS}$
must be specified consistently with Eq.(\ref{ansatz3}), for some choice of the parameters that Eq.(\ref{ansatz3}) leaves to be determined.

A natural first  guess is that the three-particle potential $V^{ABC}_{dS}$
be given (see
Eq(\ref{trivialthree}))
by a combination
of our two-particle potentials given in
Eq.(\ref{pdSsep}),
{\it i.e.}
$V^{ABC}_{dS}=V^{AB}_{dS}+V^{BC}_{dS}+V^{AC}_{dS}$, but one can easily check that this does not commute with the three-particle proper-dS charges \eqref{dscharge3}.
What does work is adding an extra term:
\begin{equation}
\label{Hvanillaproper}    V^{ABC}_{dS}=V^{AB}_{dS}+V^{BC}_{dS}+V^{AC}_{dS}
+V^{ABC}_{dS(\bigstar)}
\end{equation}
with
\begin{equation}
\label{VdSNS}
    \begin{aligned}
V^{ABC}_{dS(\bigstar)}=&\frac{g\ell^2}{2}\Bigl(p_1^Cp_2^B(x_1^Cx_2^A-x_2^Cx_1^A)-p_1^Cp_2^Ax_2^Cx_1^B-p_2^Cp_1^B(x_1^Cx_2^A-x_2^Cx_1^A)-p_2^Cp_1^Ax_1^Cx_2^B+\\      &+p_1^Bp_1^Ax_2^C(2x_2^C-x_2^A-x_2^B)-p_1^Bp_2^Ax_2^C(2x_1^C-x_1^B)-2p_2^Bp_1^Ax_1^Cx_2^C+\\&+p_2^Bp_2^Ax_1^C(2x_1^C-x_1^A-x_1^B)+p_1^Ax_1^Cp_2^Bx_2^B+x_1^Cp_1^Cp_2^Ax_2^B+\\&+x_2^Cp_2^Cp_1^Ax_1^B+x_1^Ap_1^Ap_2^Bx_2^C+x_2^Ap_2^Ap_1^Bx_1^C\Bigr)
    \end{aligned}
\end{equation}
One can easily check that the $H^{ABC}_{dS}$ of Eqs.\eqref{Hds3},
(\ref{Hvanillaproper}),
(\ref{VdSNS}) commutes with
the proper-dS charges \eqref{dscharge3}. Most importantly
  we find that indeed our
Hamiltonian $H^{ABC}_{dS}$
{\underline{uniquely}} selects the proper-dS charges \eqref{dscharge3}
 as its conserved charges.
In order to see this we start from a general parametrization of the three-particle charges
\begin{equation}
\begin{aligned}
\label{chargeansatz3}
   & \Tilde{P_1}^{tot}=\sum p_1^I+\ell  \Tilde{\gamma}_{i j}^{I J} p_i^I p_j^J + \ell^2 \Tilde{\Gamma}_{i j k}^{I J K} p_i^I p_j^J p_k^K\\
   &    \Tilde{ P_2}^{tot}=\sum p_2^I+\ell  \Tilde{\theta}_{i j}^{I J} p_i^I p_j^J + \ell^2  \Tilde{\Theta}_{i j k}^{I J K} p_i^I p_j^J p_k^K\\
      & \Tilde{R}^{tot} = \sum R^I+\ell  \Tilde{\phi}_{i}^{I J} p_i^I R^J + \ell^2  \Tilde{\Phi}_{i j }^{I J K} p_i^I p_j^J R^K
\end{aligned}
\end{equation}
which shares the same properties outlined for the two-particle ansatz  \eqref{chargeansatz} (the particle indices run over $\{A,B,C\}$ and $\Tilde{\gamma},\Tilde{\theta},\Tilde{\phi},\Tilde{\Gamma},\Tilde{\Theta},\Tilde{\Phi}$ are sets of real coefficients).

We find that by requesting that these charges  commute with our $H^A_K + H^B_K + V^{AB}_{dS}$
the parameters in Eq.\eqref{chargeansatz3} get fully fixed, giving indeed
the proper-dS charges \eqref{dscharge3}.

We leave to future studies the task of exploring the meaning of the extra term
$V^{ABC}_{dS(\bigstar)}$.
Whereas the potential in the original three-particle Hamiltonian
$H^{ABC}_0$ of
Eq.(\ref{trivialthree})
was just a sum of two-particle potentials,
we found that the potential in its correct ``proper-dS deformation" $H^{ABC}_{dS}$
must include the extra term
$V^{ABC}_{dS(\bigstar)}$ which is cubic in the observables of the three particles and is made of all terms involving simultaneously observables of all the three particles.

Also noteworthy is that
for the three-particle case the proper-dS composition gives charges which are not symmetric under particle exchange (see \eqref{dscharge3}) and accordingly
our Hamiltonian $H^{ABC}_{dS}$ also is
not symmetric under particle exchange. We do not see any objective problem with this lack of particle-exchange symmetry, but still it is a bit unsettling. This made us interested in investigating which charges would be conserved if we adopted a particle-exchange symmetrized version of our Hamiltonian $H^{ABC}_{dS}$
\begin{equation}
\label{HdSasso1}
H^{ABC}_{dS(sym)}=H_K^A+H_K^B+H_K^C+V^{AB}_{dS}+V^{BC}_{dS}+V^{AC}_{dS}+\frac{1}{6}\sum_{\pi(A,B,C)}
V^{\pi(ABC)}_{dS(\bigstar)}
\end{equation}
{\it i.e.} the Hamiltonian obtained by summing over all the possible particle permutations, $\pi(ABC)$,
of the extra term.

We then ask for which choices of the parameters of our Eq.\eqref{chargeansatz3} the Hamiltonian $H^{ABC}_{dS(sym)}$
commutes with the charges parametrized in our Eq.\eqref{chargeansatz3}, and we find that
$H^{ABC}_{dS(sym)}$
{\underline{uniquely}} selects  as its conserved charges the following ones
\begin{equation}
\label{Asso1}
\begin{aligned}
\mathcal{P}^{dS(sym)}_1&=\frac{1}{3}[(p^A\oplus_{dS} p^B)\oplus_{dS} p^C+p^A\oplus_{dS}(p^B\oplus_{dS} p^C)+(p^A\oplus_{dS} p^C)\oplus_{dS} p^B]_1=\\
&=p_1^A+p_1^B+p_1^C-\ell\bigl(p_1^Ap_2^B+p_1^Bp_2^A+p_1^Ap_2^C+p_1^Cp_2^A+p_1^Bp_2^C+p_1^Cp_2^B\bigr )+\\
&+\frac{\ell^2}{2}\bigl(p_1^A((p_2^B)^2+(p_2^C)^2)+(p_1^B+p_1^C)(p_2^A)^2-p_1^A(p_1^B)^2-p_1^B(p_1^A)^2+\\&-p_1^C(p_1^B)^2-p_1^B(p_1^C)^2+p_1^B(p_2^C)^2+p_1^C(p_2^B)^2+
+p_1^Cp_2^Cp_2^B+p_1^Bp_2^Bp_2^C+\\&-p_1^A(p_1^C)^2-p_1^C(p_1^A)^2-4p_1^Ap_1^Bp_1^C+\frac{8}{3}(p_1^Ap_2^Bp_2^C+p_1^Bp_2^Ap_2^C+p_1^Cp_2^Bp_2^A)+\\
&+p_2^A(p_1^Cp_2^C+p_1^Bp_2^C+p_1^Ap_2^C+p_1^Ap_2^B)\bigr)\\ \\
\mathcal{P}^{dS(sym)}_2&=\frac{1}{3}[(p^A\oplus_{dS} p^B)\oplus_{dS} p^C+p^A\oplus_{dS}(p^B\oplus_{dS} p^C)+(p^A\oplus_{dS} p^C)\oplus_{dS} p^B]_2= \\
&=p_2^A+p_2^B+p_2^C+\ell (p_1^C (p_1^B+p_1^A)+p_1^B p_1^A)+\\
&-\frac{1}{6} \ell^2 (3 (p_1^C)^2 (p_2^B+p_2^A)-p_1^C (3 p_2^C (p_1^B+p_1^A)+3 p_1^B p_2^B-4 p_1^B p_2^A-4 p_2^B p_1^A+3 p_1^A p_2^A)+\\
&+p_2^C (3 (p_1^B)^2+4 p_1^B p_1^A+3 (p_1^A)^2)+3 (p_1^B-p_1^A) (p_1^B p_2^A-p_2^B p_1^A))\\ \\
\mathcal{R}^{dS(sym)}&=R^A+R^B+R^C
\end{aligned}
\end{equation}
which are indeed symmetric under particle exchange. Moreover, these charges
$\mathcal{P}^{dS(sym)}_1$,
$\mathcal{P}^{dS(sym)}_2$,
$\mathcal{R}^{dS(sym)}$ close the
algebra \eqref{spatialalg}.

\section{Charges with $\kappa$-coproduct composition}
\label{kvanilla}
We now move on to applying the same strategy of the analysis to the
coproduct composition law of Eqs.\eqref{opm}-\eqref{opr}, which we rewrite here (at order $\ell^2$) for convenience
\begin{equation}
\label{vanillacharges}
\begin{aligned}
&\mathcal{P}_1=(p^A\oplus_\kappa p^B)_1=p^A_1+p^B_1-\ell p_2^Ap_1^B+\frac{\ell^2}{2}(p_2^A)^2p_1^B\\
&\mathcal{P}_2=(p^A\oplus_\kappa p^B)_2=p^A_2+p^B_2\\
&\mathcal{R}=R^A\oplus_\kappa R^B=R^A+R^B-\ell p_2^A R^B+\frac{\ell^2}{2}(p_2^A)^2R^B \, ,
\end{aligned}
\end{equation}
 As done for the proper-dS case, our first objective is to find a Hamiltonian
$H^{AB}_{\kappa}$, deformation of the  $H^{AB}_0$
of Eq.(\ref{h0ab}), such that $[\Vec{\mathcal{P}},
H^{AB}_{\kappa}]=0$ and $[\mathcal{R},
H^{AB}_{\kappa}]=0$.
Applying the same strategy of the previous section, we find that the Hamiltonian $H_\kappa^{AB}=H_K^A+H_K^B+V^{AB}_\kappa$ with
\begin{equation}
    \label{Hvanilla}
    \begin{aligned}
   & V^{AB}_\kappa=\frac{g}{2}\Big[(\Vec{x}^A-\Vec{x}^B)^2+\\
    & 2\ell \qty(-p_2^A(x_1^A)^2+x_1^Ap_2^Ax_1^B+\frac{1}{2}x_2^Ap_1^Ax_1^A+\frac{1}{2}x_2^Ax_1^Ap_1^A-x_2^Bp_1^Ax_1^A-\frac{1}{2}x_2^Bp_1^Ax_1^A)\\
&2\ell^2\qty((p_2^A)^2(x_1^A)^2+\frac{1}{2}x_1^A(p_1^A)^2x_1^A-\frac{1}{2}x_1^A(p_2^A)^2x_1^B
    )\Big]
    \end{aligned}
\end{equation}
is such that indeed
$[\Vec{\mathcal{P}},H_K^A+H_K^B+V^{AB}_\kappa] =0$ and $[\mathcal{R},H_K^A+H_K^B+V_\kappa^{AB}]=0$.
And we find that
the
Hamiltonian
$H^A_K + H^B_K + V^{AB}_{\kappa}$
{\underline{uniquely}} selects the $\kappa$-coproduct charges (\ref{vanillacharges})
 as its conserved charges.
 This is easily shown by starting again from the general  charge ansatz \eqref{chargeansatz} and requiring that they  commute with $H_K^A+H_K^B+V_{\kappa}^{AB}$:
this requirement fully fixes all the parameters in Eq.\eqref{chargeansatz}, giving indeed the
$\kappa$-coproduct charges (\ref{vanillacharges}).

It is noteworthy that the
$\kappa$-coproduct charges (\ref{vanillacharges})
are not symmetric under the exchange of particles $A$ and $B$,
and accordingly also our Hamiltonian $H_\kappa^{AB}$
is not symmetric (because the potential $V^{AB}_\kappa$ of
\eqref{Hvanilla} is not symmetric). We found that the analogous issue of lacking particle-exchange symmetry that we encountered in our analysis of the proper-dS composition law
could be ``fixed" by resorting to a symmetrized version of the Hamiltonian, but for the $\kappa$-coproduct composition law this is not the case: if one considers the symmetrized Hamiltonian \begin{equation}
\label{Hsymm1}
H_{\kappa(sym)}^{AB}=\frac{H^{AB}_\kappa+H^{BA}_\kappa}{2}
\end{equation}
then one finds that no choice of the parameters in  \eqref{chargeansatz} leads to charges that commute with $H^{\kappa(sym)}_{AB}$.

For the three-particle case the $\kappa$-coproduct composition law gives
\begin{equation}
\label{vanillacharge3}
\begin{aligned}
\tilde{\mathcal{P}_1}=(p^A\oplus_{\kappa} p^B\oplus_{\kappa}p^C)_1&=p_1^A+p_1^B+p_1^C+\ell\left(-p_1^Bp_2^A-p_1^C(p_2^A+p_2^B)\right)+\\
&+\ell^2\left(\frac{p_1^B(p_2^A)^2}{2}+\frac{1}{2}p_1^C(p_2^A+p_2^B)^2\right)\\ \\
\tilde{\mathcal{P}_2}=(p^A\oplus_{\kappa} p^B\oplus_{\kappa}p^C)_2&=p_2^A+p_2^B+p_2^C\\ \\
\tilde{\mathcal{R}}=(R^A\oplus_{\kappa} R^B\oplus_{\kappa}R^C)&=R^A+R^B+R^C+\ell \left(-p_2^A \left(R^B\right)-R^C \left(p_2^A+p_2^B\right)\right)+\\
&+\ell^2 \left(\frac{1}{2} R^C \left(p_2^A+p_2^B\right){}^2+\frac{1}{2} (p_2^{A})^2 R^B\right)
\end{aligned}
\end{equation}
Using the same procedure of \Cref{pdS} one finds that the Hamiltonian
\begin{equation}
\label{Hcoproduct3}
H_\kappa^{ABC}=H_K^A+H_K^B+H_K^C+V_\kappa^{AB}+V_\kappa^{BC}+V_\kappa^{AC}+V^{ABC}_{\kappa(\bigstar)} \, ,\end{equation}
with
\begin{equation}
    \label{Hvanilla3}
    \begin{aligned}
V^{ABC}_{\kappa(\bigstar)}=&g\ell\qty(p_1^B(x_1^Cx_2^A-x_2^Cx_1^A+x_1^Ax_2^B-x_1^Cx_2^B)+p_2^Bx_1^B(x_1^C-x_1^A))+\\
       +&g\frac{\ell^2}{2}\bigl((p_1^B)^2(x_1^Cx_1^A-x_1^Bx_1^C+x_1^Bx_1^A)-(p_2^B)^2(x_1^Cx_1^A-2x_1^Ax_1^B)+p_1^Bx_1^C(p_1^Ax_1^A-p_2^Bx_2^B)+\\
       +& p_1^Bp_2^A(x_2^Cx_1^A-x_1^Ax_2^B)+p_2^B(p_1^Bx_1^Ax_2^C+p_2^Ax_1^Ax_1^B)\bigr)
\, ,    \end{aligned}
\end{equation}
commutes with $\Vec{\Tilde{{\mathcal{P}}}}$ and $\Tilde{\mathcal{R}}$.
It is noteworthy that the $\kappa$-coproduct extra term $V^{ABC}_{\kappa(\bigstar)}$, besides involving   terms that depend  simultaneously on observables of all three particles, also involves terms that depend only on two of the particles (and these additional terms cannot be re-absorbed in a redefinition of the potentials $\Tilde{V}^{IJ}_\kappa$
since they are different for different pairs of particles).

Also in this case we find that the
Hamiltonian $H_\kappa^{ABC}$ of our Eq.(\ref{Hcoproduct3})
{\underline{uniquely}} selects the $\kappa$-coproduct charges (\ref{vanillacharge3})
 as its conserved charges:
 by requesting that the parametrized charges  of Eq.\eqref{chargeansatz3} commute with $H_\kappa^{ABC}$
the parameters in Eq.\eqref{chargeansatz3} get fully fixed, giving indeed
the $\kappa$-coproduct charges (\ref{vanillacharge3}).

$H_\kappa^{ABC}$
is not symmetric under particle exchange and its symmetrized version,
\begin{equation}
\label{Hkappasimm3}
H_{\kappa(sym)}^{ABC}=H_K^A+H_K^B+H_K^C+\frac{1}{6}\sum_{\pi(A,B,C)}
V^{\pi(ABC)}_{\kappa} \,\, ,
\end{equation}
is not a viable alternative since it does not have any conserved charges:
there is no choice of
the parameters in Eq.\eqref{chargeansatz3}
such that
the
parametrized charges  of Eq.\eqref{chargeansatz3} commute with $H_{\kappa(sym)}^{ABC}$.

\section{Closing remarks}

Inevitably, the physics community is approaching the challenge of understanding the deformed relativistic symmetries of some quantum spacetimes from a perspective which is mainly informed by our experience with special relativity, but a price can be payed when we unknowingly make inferences based on the linearity of most special-relativistic laws. In particular, the way in which special relativity governs how free-particle charges combine in conservation laws applicable when particles
interact is completely governed by the linearity of transformation laws, so that charges inevitably combine linearly.
Working within special relativity one does not even fully appreciate how the chosen form of interaction could affect the conservation laws, because the linearity of transformation laws imposes that in all cases charges combine linearly, independently of the type of interactions being considered. This is probably the reasons why, before the study we here reported, the debate on total charges for quantum spacetimes had not contemplated a possible role for the interactions, and instead relied  on one or
 another ``naturalness argument" based on the form of the relativistic properties of free particles.

We here showed, using the toy model of spatial 2D $\kappa$-Minkowski, that the nonlinearity of deformed-relativistic transformation laws is such that the correct notion of total charge depends strongly on how one introduces interactions among particles. We found that, starting from the same description of free particles, for interacting particles one can have at least three different ways for obtaining total charges:
the one based on
the proper-dS composition law, the one based on
the $\kappa$-coproduct composition law,
and the one obtained by symmetrizing
the proper-dS composition law. Interestingly, we also found that it is instead not possible to introduce interactions such that conservation laws are
obtained by symmetrizing
the $\kappa$-coproduct composition law.

We are confident that the lessons learned within the spatial 2D $\kappa$-Minkowski toy model apply also to other quantum spacetimes. Where we suspect that the specific form
of spacetime quantization might affect the analysis is the required level of complexity of Hamiltonians.
The Hamiltonians we here exhibited, the ones that do enjoy deformed relativistc invariance, are not very simple. To the human eye they appear to be unpleasantly complex, and it would be surprising (though of course possible) that Nature would choose such complex ways to describe interactions among particles. It is therefore natural to wonder if some ways to quantize spacetime with deformed relativistic symmetries could
produce simpler descriptions of interactions among particles. If such an aspect of simplicity was found for a certain scheme of spacetime quantization it might provide encoragement for the studies of other aspects of that quantum spacetime.

\section*{Acknowledgements}
We are grateful for financial support by the Programme STAR Plus, funded by Federico II University and Compagnia di San Paolo, and by the MIUR, PRIN 2017
grant 20179ZF5KS.

\end{document}